\documentclass[12pt,reqno]{amsart}

\usepackage{amsmath,amsfonts,amsbsy,amsthm}
\usepackage{amssymb}
\usepackage{eucal}
\usepackage{hyperref}
\usepackage{dsfont}
\usepackage{xy}
\xyoption{matrix} \xyoption{arrow} \xyoption{arc} \xyoption{color}
\usepackage{enumerate}
\usepackage{enumitem}
\setlist{leftmargin=*}
\usepackage{tikz}
\usetikzlibrary{decorations.pathmorphing,shapes,arrows}
\usepackage{verbatim} 
\usepackage{hyperref}

\numberwithin{equation}{section}

\makeatletter
\newtheoremstyle{corsivo}
   {\medskipamount}{\medskipamount}%
   {\itshape}{}%
   {\bfseries}{}%
   { }
   {\thmname{#1}\thmnumber{\@ifnotempty{#1}{ }\@upn{#2}}%
    \thmnote{ {\bfseries\boldmath(#3)}}.}%
\makeatother

\theoremstyle{corsivo}
\newtheorem{theorem}{Theorem}[section]
\newtheorem{lemma}[theorem]{Lemma}
\newtheorem{corollary}[theorem]{Corollary}
\newtheorem{proposition}[theorem]{Proposition}
\newtheorem{definition}[theorem]{Definition}
\newtheorem{assumption}[theorem]{Assumption}
\makeatletter
\newtheoremstyle{dritto}
   {\medskipamount}{\medskipamount}%
   {\rmfamily}{}%
   {\bfseries}{}%
   { }
   {\thmname{#1}\thmnumber{\@ifnotempty{#1}{ }\@upn{#2}}%
    \thmnote{ {\bfseries\boldmath(#3)}}.}%
\makeatother
\theoremstyle{dritto}
\newtheorem{remark}[theorem]{Remark}

\pdfoutput=1

\newcommand{\eps}{\varepsilon}
\newcommand{\Id}{\mathds{1}}  
\newcommand{\iu}{\mathrm{i}} 
\newcommand{\di}{\mathrm{d}}
\newcommand{\N}{\mathbb{N}_0}
\newcommand{\Np}{\mathbb{N}}
\newcommand{\Z}{\mathbb{Z}}
\newcommand{\R}{\mathbb{R}}
\newcommand{\Do}{\mathcal{D}}
\newcommand{\Hi}{\mathcal{H}}
\newcommand{\norm}[1]{\left\| #1 \right\|}
\newcommand{\ie}{{\sl i.\,e.\ }}   
\newcommand{\eg}{{\sl e.\,g.\ }} 
\newcommand{\cf}{{\sl cf.\ }}  
\newcommand{\virg}[1]{``#1''}
\newcommand{\half}{\mbox{\footnotesize $\frac{1}{2}$}}
\renewcommand{\(}{\left(}
\renewcommand{\)}{\right)}
\newcommand{\E}{{\mathrm{e}}}
\newcommand{\Or}{{\mathcal{O}}}
\newcommand{\V}[1]{\mathbf{#1}}
\newcommand{\abs}[1]{\left\lvert#1\right\rvert}
\newcommand{\sub}[1]{_{\mathrm{#1}}}
\newcommand{\subm}[2]{_{\mathrm{#1},#2 }}

\DeclareMathOperator{\Tr}{Tr}         

\let\oldfootnote\footnote
\renewcommand{\footnote}[1]{\oldfootnote{\  #1}}

\title[Improved energy estimates]{Improved energy estimates for a class of time-dependent perturbed Hamiltonians}
\author[G. Marcelli]{Giovanna Marcelli}

\begin{document}

\begin{abstract} 
We consider time-dependent perturbations which are relatively\\ bounded with respect to the square root of an unperturbed Hamiltonian operator, and whose commutator with the latter is controlled by the full perturbed Hamiltonian. The perturbation is modulated by two auxiliary parameters, one regulates its intensity as a prefactor and the other one controls its time-scale via a regular function, whose derivative is compactly supported in a finite interval. 
We introduce a natural generalization of energy conservation in the case of time-dependent Hamiltonians:
the boundedness of the two-parameter unitary propagator for the physical evolution with respect to the \emph{$n/2$-th power energy norm} for all $n\in\Z$.
We provide bounds of the $n/2$-th power energy norms, uniformly in time and in the time-scale parameter, for the unitary propagators, generated by the time-dependent perturbed Hamiltonian and by the unperturbed Hamiltonian in the interaction picture.
The physically interesting model of Landau-type Hamiltonians with an additional weak and time-slowly-varying electric potential of unit drop is included in this framework.

\medskip

\end{abstract}

\maketitle

\vspace{-12mm}
\tableofcontents

\newpage
\goodbreak

\section{Introduction}
We consider the physical evolution of a quantum system in a separable Hilbert space $\Hi$ generated by the time-dependent Hamiltonian operator
\begin{equation}
\label{Ch6eqn:defn H pertubed}
H(\eps,\eta, t):=H_0+\eps g(\eta t)H_1\quad\text{for all $t\in\R$},
\end{equation}
where $H_0$ is the unperturbed Hamiltonian, $H_1$ is the perturbation switched on by a function $g$ with $\mathrm{supp}\,g' \subset (0,1)$ and $g(s)=0$ for $s<0$, and $\eps\in (0,\eps_*]$, $\eta>0$ are parameters\footnote{The value $\eps_*$ will be fixed by inequality~\eqref{eqn:cond on eps star} in order to guarantee a uniform positive lower bound, precisely $1$, for $H(\eps,\eta, t)$ (see condition~\eqref{Ch6eqn:cond on inf spect}).} regulating respectively the \emph{intensity} and the \emph{time-scale} of the perturbation. The variable $t$ here stands for time and the positive parameter $\eta$ is a convenient tool to control the rate at which the system changes. The function $g$ regulates the switch-on time of the \emph{external} Hamiltonian $\eps H_1$ (notice that the perturbation is completely off for $t\leq0$).

When the Hamiltonian $H(\eps,\eta,t)$ is $t$-independent\footnote{In this case the $\eta$-dependence plays no role, thus we cancel it.}, namely $H(\eps,\eta,t)=H(\eps)$, it is well known that, by an elementary consequence of Stone's theorem, one has that $[U_\eps (t),H(\eps)]=0$, where $U_\eps(t)$ denotes the unitary propagator for the self-adjoint operator $H(\eps)$. In other words there is conservation of the energy and consequently one obtains that $H^{-n/2}(\eps)U_\eps(t) H^{n/2}(\eps)$ has a bounded extension for every $n\in\Z$. 
On the other hand, if there is a non-trivial $t$-dependence and the perturbation commutes with the unperturbed Hamiltonian, \ie $[H_1,H_0]=0$, to establish that for all $n\in\Z$ the product $H^{-n/2}(\eps,\eta, t)U_{\eps,\eta} (t,r) H^{n/2}(\eps,\eta, r)$ extends to a bounded operator, one can use the representation formula for the unitary propagator $U_{\eps,\eta} (t,r)=\E^{-\iu\int_r^t \di s\, H(\eps,\eta, s)}$ (see \cite[Proposition 2.5]{NickelSchnaubeltChapter6}) and rely on similar techniques developed in Proposition~\ref{Ch6lem:bound prod opposite power H and H_eps}.
In this paper, we deal with the more general case in which the commutator $[H_1,H_0]\neq 0$ and \virg{is controlled} by the full perturbed Hamiltonian $H(\eps,\eta,t)$, uniformly in $(\eps,\eta,t)$ (see Assumption~\ref{Ch6item:[H_1,H(eps,t)] H^-1 bounded}), beyond Assumption \ref{Ch6item:H_1 H_0^1/2 bounded} on the perturbation $H_1$ to be self-adjoint and relatively bounded with respect to $H_0^{1/2}$ (see the hypotheses in the statement of Theorem \ref{Ch6prop:en bounded}).

Unlike for time-independent Hamiltonians there is no immediate notion of energy conservation, but the boundedness of the unitary propagator for the physical evolution with respect to \emph{$n/2$-th power energy norm} arises as a natural generalization for time-dependent Hamiltonians.
Specifically, fix $n\in\Np$, defining the $n/2$-th power energy norm $\norm{\,\cdot\,}_{H^{n/2}(\eps,\eta, t)}$ of $H(\eps,\eta, t)$ as the graph norm of $H^{n/2}(\eps,\eta, t)$, namely
\[
\norm{\psi}_{H^{n/2}(\eps,\eta, t)}:=\norm{\psi}+\norm{H^{n/2}(\eps,\eta, t)\psi}\quad\text{for any }\psi\in\Do(H^{n/2}(\eps,\eta, t))
\]
and equipping $\Do(H^{n/2}(\eps,\eta, t))$ with $\norm{\,\cdot\,}_{H^{n/2}(\eps,\eta, t)}$, we introduce the space 
\[
\mathcal{L}^{(n)}_{\eps,\eta}(r,t):=\{\text{$A\colon\!\Do\(H^{n/2}(\eps,\eta, r)\)\!\to\!\Do\(H^{n/2}(\eps,\eta, t)\)$ linear and bounded}\}.
\]
Denoting by $U_{\eps,\eta}(t,r)$ the unitary propagator generated by $H(\eps,\eta,t)$, we will prove that for every $n\in\Np$ one has that $U_{\eps,\eta}(t,r)$ is in $\mathcal{L}^{(n)}_{\eps,\eta}(r,t)$ with the corresponding operator norm $\norm{U_{\eps,\eta}(t,r)}_{\mathcal{L}^{(n)}_{\eps,\eta}(r,t)}$ uniformly bounded in the parameters $(\eta,(t, r))\in (0,\infty)\times\R^2$, which is equivalent to establish the following estimate\footnote{We will prove this equivalent statement.}: For every $n\in\Z$, for all $\eps\in (0,\eps_*]$ and $\eta>0$ we have that
\begin{equation}
\label{eqn:en est for H}
\sup_{t,r\in\R}\,\,\sup_{\psi\in \Do(H^{n/2}(\eps,\eta,r)):\norm{\psi}=1}\norm{H^{-n/2}(\eps,\eta, t)U_{\eps,\eta}(t,r) H^{n/2}(\eps,\eta,r)\psi}\leq C_n(\eps),
\end{equation}
where the finite constant $C_n(\eps)$ is $\eta$-independent. The precise assumptions and result are stated in Theorem~\ref{Ch6prop:en bounded}. To the best knowledge of the author, in the standard results of well-posedness of non-autonomous linear evolution equations not even the statement $U(t,r)\in\mathcal{L}^{(2)}_{\eps,\eta}(r,t)$ is shown, the only exception being \cite[Theorem 5.1]{Kato70}.

Moreover, we are interested in working in the so-called \emph{interaction} or \emph{intermediate picture}\footnote{Usually, the interaction picture is performed using the unitary propagator induced by the time-independent part of the time-dependent perturbed Hamiltonian (\eg see \cite[\S X.12]{ReedSimonChapter6}). More generally, one can introduce the interaction picture via the two-parameter family of unitary operators generated by time-dependent part (see \cite[\S VIII.14 ]{MessiahChapter6}), fixing an initial time. In our framework, we choose the second kind of interaction picture with initial time $t_0=0$ .} : First one computes the unitary propagator $G(t,0)=\E^{-\iu\frac{\eps}{\eta} \phi(\eta t )H_1}$, with $\phi(s):=\int_0^s\di u\, g(u)$, generated by $\eps g(\eta t)H_1$ (\eg using again \cite[Proposition 2.5]{NickelSchnaubeltChapter6}) and then one considers the time-dependent unitarily transformed\footnote{In Section~\ref{Ch6sec:appl}, where we deal with the physically interesting model of Landau-type Hamiltonians, this unitary transformation is the gauge transformation $G(t,0)=\E^{-\iu\frac{\eps}{\eta} \phi(\eta t )\Lambda_1}$, where $H_1:=\Lambda_1$ models an electric potential of negative unit drop for an electric field pointing in the negative $1$-st direction (see Definition~\ref{Ch6defn:switch funct}).} Hamiltonian $G(t,0)^* H_0 \,G(t,0)=\E^{\iu\frac{\eps}{\eta} \phi(\eta t )H_1}H_0\E^{-\iu\frac{\eps}{\eta} \phi(\eta t )H_1}$. Setting the \emph{scaled time} or \emph{macroscopic time} $s:=\eta t$, we introduce
\begin{equation}
\label{Ch6eqn:defn H hat}
\hat{H}(\eps,\eta,s):=\E^{\iu \frac{\eps}{\eta}\phi(s)H_1}H_0\E^{-\iu \frac{\eps}{\eta} \phi(s)H_1}.
\end{equation}
Similarly to the previous case, we will prove the following inequality: For every $n\in \Z$, for all $\eps\in (0,\eps_*]$ and $\eta>0$ we have that
\begin{equation}
\label{eqn:en est for H hat}
\sup_{s,u\in\R}\,\,\sup_{\psi\in \Do\(\hat{H}^{n/2}(\eps,\eta,r)\):\norm{\psi}=1}\norm{\hat{H}^{-n/2}(\eps,\eta,s)\hat{U}_{\eps,\eta}(s,u) \hat{H}^{n/2}(\eps,\eta,u)\psi}\leq C_n(\eps) (1+\eps D_n),
\end{equation}
where $\hat{U}_{\eps,\eta}(s,u)$ is the unitary propagator generated by $\hat{H}(\eps,\eta,s)$ and $D_n$ is a finite constant independent of $(\eps,\eta)$. This result, formulated in Corollary~\ref{Ch6thm:en bounded scaled H}, is obtained as a consequence of estimate~\eqref{eqn:en est for H}, thanks to the following identity
\begin{equation}
\label{eqn:unitaries are related by the gauge transf}
\hat{U}_{\eps,\eta}(s,u)\equiv\E^{\iu \frac{\eps}{\eta}\phi(s)H_1}U_{\eps,\eta}(s/\eta,u/\eta)\E^{-\iu\frac{\eps}{\eta} \phi(u)H_1},
\end{equation}
and Proposition~\ref{Ch6lem:bound prod opposite power H and H_eps}, which guarantees that for every integer number $n$, $H_0^{n/2}H^{-n/2}(\eps,\eta,t)$ and $H^{n/2}(\eps,\eta,t)H_0^{-n/2}$ are bounded in the operator norm by $\Or(\eps)+1$, uniformly in $(\eta,t)\in(0,\infty)\times\R$. 

Energy estimates in the form of \eqref{eqn:en est for H hat} (or equivalently \eqref{eqn:en est for H}) are relevant when one needs to keep track of localization in energy under the physical evolution, uniformly in the time-scale of the perturbation. More precisely, suppose that a family of operators $O(s)$ with $s\in \R$ \emph{decays in energy with power $m/2$} with $m\in\Np$, in the sense that there exists a finite constant $C_O$ such that 
\begin{equation}
\label{eqn: enloc}
\norm{O(s) \hat{H}^{m/2}(\eps,\eta,s)\psi}\leq C_O\norm{\psi}
\end{equation}
for every $\eps\in (0,\eps_*]$, $\eta>0$, $s\in\R$ and for all $\psi\in \Do\(\hat{H}^{m/2}(\eps,\eta,s)\)$.
Then, by applying inequality \eqref{eqn:en est for H hat} this energy localization is conserved by the evolved family of operators $\hat{U}_{\eps,\eta}(u,s) O(s) \hat{U}_{\eps,\eta}(s,u)$:
\begin{align}
\label{eqn: enlocevol}
\norm{\hat{U}_{\eps,\eta}(u,s) O(s) \hat{U}_{\eps,\eta}(s,u) \hat{H}^{m/2}(u)\psi}&=\norm{O(s)\hat{H}^{m/2}(s)\hat{H}^{-m/2}(s) \hat{U}_{\eps,\eta}(s,u) \hat{H}^{m/2}(u)\psi}\cr
&\leq C_O C_m(\eps) (1+\eps D_m)\norm{\psi},
\end{align}
for any $s,u\in\R$ and for every $\psi\in \Do\(\hat{H}^{m/2}(\eps,\eta,u)\)$.

This work has been motivated in the first instance by the need to fill a gap in the proof of \cite[Lemma 5.1]{ElgartSchlein04Chapter6}, where Landau-type Hamiltonian operators with an additional weak and time-slowly-varying electric potential of unit drop are considered (see Section~\ref{Ch6sec:appl} for this application). While Theorem~\ref{Ch6prop:en bounded} implies \cite[Lemma 5.1]{ElgartSchlein04Chapter6}, Corollary~\ref{Ch6thm:en bounded scaled H} is relevant since it is explicitly used in the proof of \cite[Theorem 2.2]{ElgartSchlein04Chapter6} (see \cite[Remark (3), p. 599]{ElgartSchlein04Chapter6} for the case $n=0$). The strategy proof of Theorem~\ref{Ch6prop:en bounded} is based on the one given in the aforementioned paper, with two essential differences: firstly we use $H(\eps,\eta,t)$ whose time derivative is compactly supported (while $\frac{\partial}{\partial s}\hat{H}(\eps,\eta,s)$ is not compactly supported) and secondly in the proof of Theorem~\ref{Ch6prop:en bounded} we establish the induction step by computing the time derivative of the bounded operator $H^{-1/2}(\eps,\eta,t)$ (compare \eqref{Ch6eqn:en bound n=N step I}) instead of the unbounded one $\hat{H}^{1/2}(\eps,\eta,s)$. As it is briefly explained in Section \ref{Ch6sec:appl}, these kinds of energy estimates are used to prove the validity of the Kubo formula for the transverse conductance in the quantum Hall effect in a two-dimensional sample (\eg see \cite{Graf07, BachmannDeRoeckFraas17, GiulianiMastropietroPorta17, MaPaTa, Teufel20, HenheikTeufel21, MaPaTe, MarcelliMonaco1}). But we are convinced that our results are of general conceptual interest, since we provide bounds on the growth of the $n/2$-th power energy norms for time-dependent Hamiltonian in a model-independent setting, and could be relevant for proving the linear response in quantum Hall systems for unbounded Hamiltonians (\cf Section \ref{Ch6sec:appl}).
More specifically, we require mild properties: Beyond the technical hypotheses, \ie Assumptions~\ref{Ch6ass:perturbed H(t)} and \ref{domain power H0 invariant H1} for $k=2$, which guarantee the self-adjointness of  $H(\eps,\eta, t)$ and $\hat{H}(\eps,\eta,s)$ on the same $t$-independent domain $\Do(H_0)$ and spectrum condition~\eqref{Ch6eqn:cond on inf spect}, the operator $H_1$ associated with the perturbation must not be bounded but only $H_0^{1/2}$-bounded (compare Assumption~\ref{Ch6item:H_1 H_0^1/2 bounded}), and the two parameters $\eps,\eta$, related to the perturbation, are independent. Furthermore, both estimates \eqref{eqn:en est for H} and \eqref{eqn:en est for H hat} are uniform in the time-scale parameter $\eta>0$, while for fixed $\eta>0$ these bounds are clearly expected, due to the hypothesis $\mathrm{supp}\,g' \subset (0,1)$, with $\eta$-dependent constants.
Finally, the use of the symbols $\eps$ and $\eta$ is not related to a smallness assumption, as far as this paper is concerned (however our results apply to the particular case considered in \cite{ElgartSchlein04Chapter6}, where the limit $\eps=\eta=\frac{1}{\tau}\to 0^+$ is considered).

\medskip

\noindent \textbf{Acknowledgments.}  
I would like to thank Horia Cornean, Marco Falconi, Gian Michele Graf, Gianluca Panati, Benjamin Schlein, and Stefan Teufel for useful discussions and valuable comments. G.\,M. acknowledges the financial support from the European Research Council (ERC), under the European Union's Horizon 2020 research and innovation programme (ERC Starting Grant MaMBoQ, grant agreement No. 802901).

\medskip

\noindent \textbf{Data availability statement.}  
Data sharing not applicable to this article as no datasets were generated or analysed during the current study.
\goodbreak

\section{Mathematical setting and main results}
\label{sec: settingresults}
In this section we set up the mathematical framework and state our main results, under different assumptions.
Let $\Hi$ denote a separable Hilbert space. 

\noindent 
Firstly, we write hypotheses on each summand of the perturbed Hamiltonian $H(\eps,\eta,t)$.
\begin{assumption}
\label{Ch6ass:perturbed H(t)}
Let $H(\eps,\eta,t)$ be as in \eqref{Ch6eqn:defn H pertubed} and $g \in C^k(\R)$ with\footnote{Notice that we do not require that $\mathrm{supp}g$ is compact.} $k\geq 1$, $\mathrm{supp}\,g' \subset (0,1)$ and $g(s)=0$ for $s<0$. We define 
\begin{equation}
\label{eqn:def M and M'}
M:=\max_{s\in[0,1]}\abs{g(s)}\text{ and }{M}^{\prime}:=\max_{s\in[0,1]}\abs{g' (s)}.
\end{equation}
Here $\eps\in (0, \eps_*]$, where $\eps_*$ is chosen so that condition \eqref{eqn:cond on eps star} is fulfilled, and $\eta>0$.
Furthermore, the Hamiltonian operator $H(\eps,\eta,t)$ satisfies the following properties: 
\begin{enumerate}[label={\rm (A$_\arabic*$)},ref={\rm (A$_\arabic*$)}]
\item \label{Ch6item:H_0 ass} $H_0\colon \Do(H_0)\to \Hi$ is self-adjoint, where $\Do(H_0)\subset \Hi$ denotes its dense domain, and\footnote{The following hypothesis is equivalent, up to a shift of a constant, to require that $H_0$ is bounded from below.} $H_0\geq 1+\gamma_0$, with $\gamma_0>0$.
\item \label{Ch6item:H_1 H_0^1/2 bounded} 
$H_1\colon\Do(H_1)\to\Hi$ is self-adjoint, where $\Do(H_1)\subset \Hi$ denotes its dense domain, and is $H_0^{1/2}$-bounded, namely there exists a finite constant $a>0$ such that $\norm{H_1 H_0^{-1/2}}\leq a$.
\end{enumerate} 
\end{assumption}
\noindent 
As it is explained respectively in Remark~\ref{rem:aft ass}.\ref{item1:rem aft ass} and Remark~\ref{rem:aft ass}.\ref{item2:rem aft ass}, the above assumptions ensure that $H(\eps,\eta,t)$ is self-adjoint on $\Do(H_0)$ and that $H(\eps,\eta,t)\geq 1$.

\noindent
Secondly, we write hypotheses on \virg{how the perturbed Hamiltonian $H(\eps,\eta,t)$ behaves with respect to the unperturbed one $H_0$}. 
\begin{assumption}
\label{Ch6ass:perturbed H(t)_ndip}
Let $H(\eps,\eta,t)$ be as in Assumption~\ref{Ch6ass:perturbed H(t)}.\\
For every $k\in\Z$, there exists a finite constant $E_k$ such that for all $\eps\in (0, \eps_*],\eta\in (0,\infty), t\in\R$ we have that\footnote{Notice that we are allowed to write any negative power of $H(\eps,\eta,t)$ due to condition \eqref{Ch6eqn:cond on inf spect}.}: \\
if $k\geq 0$ taking any $\psi\in \Do(H^{(k+1)/2}(\eps,\eta,t))$ otherwise $\psi\in\Hi$
\begin{enumerate}[label={\rm (B($k$))},ref={\rm (B($k$))}]
\item \label{Ch6item:[H_1,H(eps,t)] H^-1 bounded} 
\[
\norm{H^{-k/2}(\eps,\eta,t)[H(\eps,\eta,t),H_1]H^{(k-2)/2}(\eps,\eta,t)\psi}\leq E_k\norm{\psi},
\]
\end{enumerate}
where $ [H(\eps,\eta,t),H_1]$ is densely defined with $\Do([H(\eps,\eta,t),H_1])\supset \Do( H^{3/2}(\eps,\eta,t))$, and in addition if $k\leq -1$ we require that $[H(\eps,\eta,t),H_1]\colon\Do(H^{(\abs{k}+2)/2}(\eps,\eta,t)) \to \Do(H^{\abs{k}/2}(\eps,\eta,t))$.
\end{assumption}

\begin{assumption}
\label{Ch6item:scaled Ham well-def} 
Let $H(\eps,\eta,t)$ be as in Assumption~\ref{Ch6ass:perturbed H(t)}.\\
For every $k\in\Np$ with\footnote{\label{fn:power1} For $k=1$ the following identity is implied by Assumptions~\ref{Ch6item:H_0 ass} and \ref{Ch6item:H_1 H_0^1/2 bounded} (see Remark~\ref{rem:aft ass}.\ref{item1:rem aft ass}).} $k\geq 2$,
\begin{enumerate}[label={\rm (C$_\arabic*$($k$))},ref={\rm (C$_\arabic*$($k$))}]
\item \label{Ch6item:power} 
for all $\eps\in (0, \eps_*],\eta\in (0,\infty), t\in\R$ we have that $\Do(H^{k}(\eps,\eta, t))\equiv\Do(H_0^k)$.
\end{enumerate}
For every $k\in\Np$ 
\begin{enumerate}[resume, label={\rm (C$_\arabic*$($k$))},ref={\rm (C$_\arabic*$($k$))}]
\item \label{domain power H0 invariant H1} we have that the domain $\Do(H^{k/2}_0)$ is invariant under the unitary transformation ${\{\E^{\iu \lambda H_1}\}}_{\lambda\in\R}$, namely for all $\lambda\in\R$ one has that
$$\E^{\iu \lambda H_1}\colon \Do(H^{k/2}_0)\to \Do(H^{k/2}_0).$$
\end{enumerate}
\end{assumption}

\begin{remark}
\label{rem:aft ass}
Here we explain some useful consequences of the hypotheses above.
\begin{enumerate}[label=(\roman*), ref=(\roman*)]
\item \label{item1:rem aft ass} Under Assumptions \ref{Ch6item:H_0 ass} and \ref{Ch6item:H_1 H_0^1/2 bounded}, we have that $H_1$ is $H_0$-bounded, with relative bound $\tilde{a}<1$. Indeed, notice that for every $C>0$
\[
\norm{H_1{\(H_0+C\)}^{-1}}=\norm{H_1H_0^{-1/2}\cdot H_0^{1/2}{\(H_0+C\)}^{-1/2}\cdot{\(H_0+C\)}^{-1/2}}\leq \frac{a}{\sqrt{1+C}},
\]
where $a$ is defined in Assumption \ref{Ch6item:H_1 H_0^1/2 bounded}. Hence, for every $\psi\in\Do(H_0)$ we obtain that
\[
\norm{H_1\psi}=\norm{H_1{\(H_0+C\)}^{-1}{\(H_0+C\)}\psi}\leq \frac{a}{\sqrt{1+C}}\( \norm{H_0\psi}+C\norm{\psi}\).
\]
Therefore, by the Kato--Rellich theorem $H(\eps,\eta,t)$ is self-adjoint on $\Do(H_0)$.
\item \label{item2:rem aft ass} Observe that Assumptions~\ref{Ch6item:H_0 ass} and \ref{Ch6item:H_1 H_0^1/2 bounded} imply that there exists $\eps_*>0$ such that
\begin{equation}
\label{Ch6eqn:cond on inf spect}
\inf_{t\in\R,\eta>0}\sigma (H(\eps,\eta,t))\geq 1\quad\text{for all $\eps\in (0, \eps_*]$}.
\end{equation}
In fact, for any $z<1$, $H(\eps,\eta,t)-z=\(\Id+ \eps g(\eta t) H_1{\(H_0-z\)}^{-1}\)\(H_0-z\)$ is invertible for a suitable choice of $\eps_*$. In view of hypothesis \ref{Ch6item:H_0 ass} and the previous remark, we get that
\begin{align*}
\norm{H_1{\(H_0-z\)}^{-1}}\leq \norm{H_1{H_0}^{-1}}\( 1+\frac{\abs{z}}{1+\gamma_0-z}\)\leq \frac{3\gamma_0+1}{\gamma_0}\norm{H_1{H_0}^{-1}}
\end{align*}
and thus there exists $\eps_*>0$ such that 
\begin{equation}
\label{eqn:cond on eps star}
\frac{3\gamma_0+1}{\gamma_0}\eps_* M\norm{H_1{H_0}^{-1}}<1
\end{equation}
 with $M$ defined in \eqref{eqn:def M and M'}.
\item \label{item3:rem aft ass} For $k\in\Np$ with $k\geq 2$, Assumption~\ref{Ch6item:power} and \cite[Supplementary notes, V.7]{KatoChapter6} imply that for all $\eps\in (0, \eps_*],\eta\in (0,\infty), t\in\R$ one has that $\Do(H^{k/2}(\eps,\eta, t))\equiv\Do(H_0^{k/2})$. The same result holds true automatically for $k=1$ due to $\Do(H(\eps,\eta, t))\equiv\Do(H_0)$ by Remark \ref{rem:aft ass}.\ref{item1:rem aft ass}.
\end{enumerate}
\end{remark}

\noindent 
Before stating the main results, namely Theorem~\ref{Ch6prop:en bounded} and Corollary~\ref{Ch6thm:en bounded scaled H}, it is convenient to recall the problem of well-posedness of non-autonomous linear evolution equations.
As it is emphasized in \cite[Notes of Section X.12]{ReedSimonChapter6}, the Cauchy problem for linear evolution equations 
\begin{equation*}
\frac{\di \psi}{\di t}(t)=A(t)\psi(t),\text{ $0\leq t\leq T$, in a Banach space}
\end{equation*}
where $A(\,\cdot\,)$ is an unbounded-operator valued function and the domain $\Do(A(t))\equiv \Do$ of $A(t)$ is independent of $t$, under general suitable conditions, was solved first by T. Kato in \cite{Kato53} and then by K. Yosida in \cite{Yosida68} (for the comparison of these works see \cite{SchmidGriesemer14}). For more general results, considering that $A(t)$ has domain which does depend on time, see \eg  \cite{Kato70, Yajima11, Schmid15} and references therein. In the present setting, under Assumption~\ref{Ch6ass:perturbed H(t)} one has that the domain of self-adjointness $\Do(H(\eps,\eta,t))$ of $H(\eps,\eta,t)$ is independent of $t$ by Remark~\ref{rem:aft ass}.\ref{item1:rem aft ass}. 
Hence, under additional hypotheses (\eg assumptions in \cite[Theorem 3]{Kato53}) one can prove that there exists the unitary propagator $U_{\eps,\eta}(t,r)$ generated by $H(\eps,\eta,t)$. This means that $U_{\eps,\eta}(t,r)$ is the two-parameter family of unitary operators, jointly strongly continuous in $t\in\R$ and $r\in\R$, such that for every  $t,r,u\in\R$
\begin{equation*}
\begin{aligned}
U_{\eps,\eta}(t,r)U_{\eps,\eta}(r,u)&=U_{\eps,\eta}(t,u),\quad U_{\eps,\eta}(t,t)=\Id,\quad U_{\eps,\eta}(t,u)\Do(H_0)=\Do(H_0),\\
\iu  \frac{\partial U_{\eps,\eta}}{\partial t}(t,u)\psi&=H(\eps,\eta,t)U_{\eps,\eta}(t,u)\psi\quad\text{ for all $\psi\in\Do(H_0)$,}\\
-\iu \frac{\partial U_{\eps,\eta}}{\partial u}(t,u)\psi&=U_{\eps,\eta}(t,u)H(\eps,\eta,u)\psi\quad\text{ for all $\psi\in\Do(H_0)$.}
\end{aligned}
\end{equation*}

\noindent
In order to keep the reader's attention on the main results, \ie Theorem \ref{Ch6prop:en bounded} and Corollary \ref{Ch6thm:en bounded scaled H}, we postpone their proofs to Section~\ref{sec: proof main results}.

\begin{theorem}
\label{Ch6prop:en bounded}
Consider the Hamiltonian $H(\eps,\eta,t)=H_0+\eps g(\eta t)H_1$ satisfying Assumption~\ref{Ch6ass:perturbed H(t)} and let $U_{\eps,\eta}(t,r)$ be the unitary propagator generated by $H(\eps,\eta,t)$. Let $n\in\Z$. If $\abs{n}\geq 2$ we assume in addition Assumption~\ref{Ch6item:[H_1,H(eps,t)] H^-1 bounded} for all $0 \leq k\leq \abs{n}-2$. Then for every $\eps\in (0, \eps_*]$ we have that 
\begin{equation}
\label{eqn: ineq thm}
\sup_{t,r\in\R}\,\,\sup_{\psi\in \Do(H^{n/2}(\eps,\eta,r)):\norm{\psi}=1}\norm{H^{-n/2}(\eps,\eta, t)U_{\eps,\eta}(t,r) H^{n/2}(\eps,\eta,r)\psi}\leq C_n(\eps)\quad\forall\,\eta>0,
\end{equation}
where $C_n(\eps)$ is defined iteratively as
\begin{equation}
\label{eqn:def Cn(eps)}
\begin{cases}
& C_0(\eps):=C_0=1\\
& C_n(\eps):=C_{n-1}(\eps)\E^{ C_{n-1}(\eps)(\alpha+\beta\eps+\gamma_n)\eps}\text{ for all $n\geq 1$}
\end{cases}
\end{equation}
with $\alpha,\beta$ and $\gamma_n$ finite constants defined as 
\begin{equation}
\label{eqn: defn consts}
\alpha+\eps\beta:=M'(a+\eps M a^2),\quad \gamma_1:=0 \text{ and } \gamma_n:=M'\sum_{k=0}^{n-2}E_k\text{ for $n\geq 2$,}
\end{equation}
and $C_{-n}(\eps):=C_n(\eps)$ for all $n\in\Np$.
\end{theorem}

\begin{remark}
Both in the Gell-Mann--Low \cite{GellMannLow} and the Kubo \cite{Kubo57} formula the standard choice for the switch-on procedure in time is to make use of the exponential function for the non-positive time-axis $\R_-:=(-\infty,0]$. More specifically, in our setting of reference, we replace the function $g$ with the exponential and restrict the whole real time-axis to the non-positive one, \ie one considers the time-dependent Hamiltonian operator
\[
H\sub{exp}(\eps,\eta, t):=H_0+\eps \E^{\eta t} H_1\quad\text{for all $t\in\R_-$}.
\]
Clearly, the main difference between $H\sub{exp}(\eps,\eta, t)$ and $H(\eps,\eta, t)\equiv H_g(\eps,\eta, t)$, defined in \eqref{Ch6eqn:defn H pertubed}, is that the the switch-on process acts respectively on the infinite time-interval $\R_-$ and on a finite time-interval (precisely, under Assumption \ref{Ch6ass:perturbed H(t)}: $\mathrm{supp}\,g' \subset (0,1)$).
Under the assumptions of Theorem \ref{Ch6prop:en bounded} except for the substitution of $g$ with the exponential and the restriction to $\R_-$ (applying the  these two replacements everywhere in the nested hypotheses), a type of inequality similar to \eqref{eqn: ineq thm} still holds true. Precisely, denoting by $U\subm{exp}{\eps,\eta}(t,r)$ the unitary propagator generated by $H\sub{exp}(\eps,\eta, t)$, we have
\begin{equation}
\label{eqn: ineq expthm}
\!\sup_{t,r\in\R_-}\sup_{\psi\in \Do(H\sub{exp}^{n/2}(\eps,\eta,r)):\norm{\psi}=1}\norm{H^{-n/2}\sub{exp}(\eps,\eta, t)U\subm{exp}{\eps,\eta}(t,r) H\sub{exp}^{n/2}(\eps,\eta,r)\psi}\leq \widetilde{C}_n(\eps)\quad\forall\eta>0,
\end{equation}
where $\widetilde{C}_n(\eps)$ is defined iteratively as
\begin{equation}
\label{eqn:def tildeCn(eps)}
\begin{cases}
& \widetilde{C}_0(\eps):=\widetilde C_0=1\\
& \widetilde{C}_n(\eps):=\widetilde{C}_{n-1}(\eps)\E^{ \widetilde{C}_{n-1}(\eps)(\widetilde{\alpha}+\widetilde{\beta}\eps+\widetilde{\gamma}_n)\eps}\text{ for all $n\geq 1$}
\end{cases}
\end{equation}
with $\widetilde{\alpha}, \widetilde{\beta}$ and $\widetilde{\gamma}_n$ finite constants, and $\widetilde C_{-n}(\eps):=\widetilde C_n(\eps)$ for all $n\in\Np$.

\noindent
Here, for completeness we sketch a proof of the above statement.
We follow the argument of the proof of Theorem~\ref{Ch6prop:en bounded} in Subsection~\ref{ssect:proofthm}, excepting the restriction of the times $t,r$ to a finite time-interval depending on $\eta$ (see \eqref{Ch6eqn:reduction of the en bounded}). Similarly, defining for every $t,r\in\R_-$
\[
\widetilde C_{\eps,\eta,n}(t,r):=\sup_{\psi\in \Do(H\sub{exp}^{n/2}(\eps,\eta,r)):\norm{\psi}=1}\norm{H\sub{exp}^{-n/2}(\eps,\eta,t)U\subm{exp}{\eps,\eta}(t,r) H\sub{exp}^{n/2}(\eps,\eta,r)\psi},
\]
one arrives at the inequality
\[
\widetilde C_{\eps,\eta,N}(t,r)\leq \widetilde  C_{N-1}(\eps)\(1+(\widetilde  \alpha+\widetilde \beta\eps +\widetilde \gamma_N)\eps  \int_t^r\di \tau\, \eta\E^{\eta \tau}\widetilde C_{\eps,\eta,N}(\tau,r)\),
\]
for $-\infty<t\leq r\leq 0 $. By using Gr\"onwall's inequality and that $\int_{-\infty}^0\di \tau\,\eta \E^{\eta \tau}=1$, we conclude that 
\begin{align*}
\widetilde C_{\eps,\eta,N}(t,r)&\leq \widetilde  C_{N-1}(\eps)\E^{ \widetilde C_{N-1}(\eps)(\widetilde \alpha+\widetilde \beta\eps+\widetilde \gamma_N)\eps\int_t^r\di \tau\, \eta \E^{\eta \tau}}\cr
&\leq \widetilde C_{N-1}(\eps)\E^{ \widetilde C_{N-1}(\eps)(\widetilde \alpha+\widetilde\beta\eps+\widetilde\gamma_N)\eps}=:\widetilde C_N(\eps).
\end{align*}

Therefore, it emerges that the crucial properties of the switch-on procedure modeled by a generic function $f\colon I\to \R$, $f\in C^k(I)$ with $k\geq 1$ on a subset $I\subseteq \R$ to deduce a type of inequality in the form of \eqref{eqn: ineq thm}, which is uniform in the time-scale parameter $\eta$, is to have that both $\norm{f}_{L^\infty(I)}$ and $\norm{f '}_{L^1(I)}$ are finite.
\end{remark}

Let the scaled time $s=\eta t$, consider the unperturbed Hamiltonian in the interaction picture $\hat{H}(\eps,\eta,s)$, defined in \eqref{Ch6eqn:defn H hat}, which is self-adjoint on $\Do(H_0)$ under Assumptions \ref{Ch6ass:perturbed H(t)} and \ref{domain power H0 invariant H1} for $k=2$. Let us briefly recall the notion of the corresponding unitary propagation, whose existence and uniqueness are guaranteed again by \cite[Theorem 3]{Kato53}, under additional regularity hypotheses. Let $\hat{U}_{\eps,\eta}(s,r)$ be the unitary propagator generated by $\hat{H}(\eps,\eta,s)$, namely $\hat{U}_{\eps,\eta}(s,r)$ is the two-parameter family of unitary operators, jointly strongly continuous in $s\in\R$ and $r\in\R$, such that for every  $s,r,u\in\R$
\begin{equation}
\label{Ch6eqn: evol U hat}
\begin{aligned}
\hat{U}_{\eps,\eta}(s,r)\hat{U}_{\eps,\eta}(r,u)&=\hat{U}_{\eps,\eta}(s,u),\quad \hat{U}_{\eps,\eta}(s,s)=\Id,\quad \hat{U}_{\eps,\eta}(s,u)\Do(H_0)=\Do(H_0),\\
\iu \eta \frac{\partial \hat{U}_{\eps,\eta}}{\partial s}(s,u)\psi&=\hat{H}(\eps,\eta,s)\hat{U}_{\eps,\eta}(s,u)\psi\quad\forall \psi\in\Do(H_0),\\
-\iu \eta \frac{\partial \hat{U}_{\eps,\eta}}{\partial u}(s,u)\psi&=\hat{U}_{\eps,\eta}(s,u)\hat{H}(\eps,\eta,u)\psi\quad\forall \psi\in\Do(H_0).
\end{aligned}
\end{equation}

\begin{corollary}
\label{Ch6thm:en bounded scaled H}
Under Assumptions~\ref{Ch6ass:perturbed H(t)} and \ref{domain power H0 invariant H1} for $k=2$, consider $\hat{H}(\eps,\eta,s)=\E^{\iu \frac{\eps}{\eta}\phi(s)H_1}H_0\E^{-\iu \frac{\eps}{\eta} \phi(s)H_1}$, where $s=\eta t$ is the scaled time.  Let $\hat{U}_{\eps,\eta}(s,u)$ be the unitary propagator generated by $\hat{H}(\eps,\eta,s)$.  Let $n\in\Z$. Let Assumption~\ref{domain power H0 invariant H1} for $k=\abs{n}$ hold true. If $\abs{n}\geq 3$ we assume in addition Assumption~\ref{Ch6item:power} for all $3\leq k\leq \abs{n}$ and Assumption~\ref{Ch6item:[H_1,H(eps,t)] H^-1 bounded} for $k=0$. If $\abs{n}\geq 4$ we assume further Assumption~\ref{Ch6item:[H_1,H(eps,t)] H^-1 bounded} for all $2-\abs{n}\leq k\leq -2$. Then there exists a finite constant $D_n$ such that for every $\eps\in(0, \eps_*]$ and $\eta\in (0,\infty)$ we have that 
$$
\sup_{s,u\in\R}\,\,\sup_{\psi\in \Do\(\hat{H}^{n/2}(\eps,\eta,r)\):\norm{\psi}=1}\norm{\hat{H}^{-n/2}(\eps,\eta,s)\hat{U}_{\eps,\eta}(s,u) \hat{H}^{n/2}(\eps,\eta,u)\psi}\leq C_n(\eps) (1+\eps D_n),
$$
where $C_n(\eps)$ is defined in \eqref{eqn:def Cn(eps)}.
\end{corollary}

\noindent
Here, we state two auxiliary results whose technical proofs are deferred to Section~\ref{sec:aux res}.

\noindent
Specifically, the following lemma shows that $H_1$ is actually $H^{1/2}(\eps,\eta, t)$-bounded with a relative bound independent of the parameters $(\eta,t)\in(0,\infty)\times\R$, not only $H^{1/2}_0=H^{1/2}(\eps,\eta, r)$-bounded with $r\leq 0$ (compare Assumption~\ref{Ch6item:H_1 H_0^1/2 bounded}). 

\begin{lemma}
\label{Ch6lem:H_0^1/2 bounded implies H^1/2(t)bounded}
Let $H(\eps,\eta,t)$ be as in Assumption~\ref{Ch6ass:perturbed H(t)}. Then for every $\eps\in(0, \eps_*]$, $\eta\in (0,\infty)$ and $t\in\R$ we have that
\[
\norm{H_1 H^{-1/2}(\eps,\eta, t)}\leq a+\eps M a^2.
\]
\end{lemma}

\noindent
On the other hand, the next proposition turns out to be useful to deduce the energy estimates for the unperturbed Hamiltonian in the interaction picture $\hat{H}(\eps,\eta,s)$ from the ones for the perturbed Hamiltonian $H(\eps,\eta,t)$.
\begin{proposition}
\label{Ch6lem:bound prod opposite power H and H_eps}
Let $H(\eps,\eta,t)$ be as in Assumption~\ref{Ch6ass:perturbed H(t)}. Let $n\in\Z$. If $\abs{n}\geq 3$ we assume in addition Assumption~\ref{Ch6item:power} for all $3\leq k\leq \abs{n}$ and Assumption~\ref{Ch6item:[H_1,H(eps,t)] H^-1 bounded} for $k=0$. If $\abs{n}\geq 4$ we assume further Assumption~\ref{Ch6item:[H_1,H(eps,t)] H^-1 bounded} for all $2-\abs{n}\leq k\leq -2$. Then there exist finite constants $A_n,B_n$ such that for every $\eps\in(0, \eps_*]$, $\eta\in (0,\infty)$ and $t\in\R$:   
\begin{enumerate}[label=(\roman*), ref=(\roman*)]

\item for any $\psi\in\Do(H^{-n/2}(\eps,\eta,t))$ we have that 
\begin{equation}
\label{Ch6eqn:bound prod opposite power H and H_eps 1}
\norm{H_0^{n/2}H^{-n/2}(\eps,\eta,t)\psi}\leq (1+ A_n\eps)\norm{\psi},
\end{equation}

\item for any $\psi\in\Do(H^{-n/2}_0)$ we have that 
\begin{equation}
\label{Ch6eqn:bound prod opposite power H and H_eps 2}
\norm{H^{n/2}(\eps,\eta,t) H_0^{-n/2}\psi}\leq (1+ B_n\eps)\norm{\psi}.
\end{equation}
\end{enumerate}
\end{proposition}

\section{Proof of the main results}
\label{sec: proof main results}
\subsection{Proof of Theorem~\ref{Ch6prop:en bounded}}
\label{ssect:proofthm}
First of all, notice that it suffices to check inequality~\eqref{eqn: ineq thm} for $n\in\N$ due to the Riesz Lemma. In view of the hypothesis $\mathrm{supp}\,g' \subset (0,1)$, for any $\psi\in\Do(H_0)$ the map $t\mapsto H(\eps,\eta,t)\psi$ is time-independent for $t\leq 0$ and $t\geq 1/\eta$. Therefore, it is enough to prove that for all $n\in\N$
\begin{equation}
\label{Ch6eqn:reduction of the en bounded}
\sup_{t,r\in [0,1/\eta]}\,\,\sup_{\psi\in \Do(H^{n/2}(\eps,\eta,r)):\norm{\psi}=1}\norm{H^{-n/2}(\eps,\eta,t)U_{\eps,\eta}(t,r) H^{n/2}(\eps,\eta,r)\psi}\leq C_n(\eps).
\end{equation}
Indeed, defining 
\begin{equation}
\label{Ch6eqn:defn C_eps,n(t,r)}
C_{\eps,\eta,n}(t,r):=\sup_{\psi\in \Do(H^{n/2}(\eps,\eta,r)):\norm{\psi}=1}\norm{H^{-n/2}(\eps,\eta,t)U_{\eps,\eta}(t,r) H^{n/2}(\eps,\eta,r)\psi},
\end{equation}
we have
\begin{equation}
\label{Ch6eqn:sup split in 4 2nd line}
\sup_{t,r\in\R}C_{\eps,\eta,n}(t,r)=\sup_{t,r\in[0,1/\eta]}C_{\eps,\eta,n}(t,r).
\end{equation}
To prove the last equality it suffices to notice that for all $t\in \R$: if $r<0$ then $C_{\eps,\eta,n}(t,r)=C_{\eps,\eta,n}(t,0)$, and similarly if $r>1/\eta$ then $C_{\eps,\eta,n}(t,r)=C_{\eps,\eta,n}(t,1/\eta)$, using that $H(\eps,\eta,r)$ is constant for $r\in\R\setminus (0,1/\eta)$ and $U_{\eps,\eta}(t,r)=U_{\eps,\eta}(t,s)U_{\eps,\eta}(s,r)$ for all $t,s,r\in\R$. One obtains analogous identities exchanging the roles of $r$ and $t$.
In order to prove inequality~\eqref{Ch6eqn:reduction of the en bounded}, we proceed by induction over $n\in\N$.
For $n=0$ it is trivial. Now we take some $N\in\N$ with $N\geq 1$. We assume that the thesis holds true for $n=N-1$ and we prove it for $n=N$. Let us start by noticing that for every $\psi\in\Do(H_0)$, we have that 
\begin{align}
\label{Ch6eqn:en bound n=1 step I}
&U_{\eps,\eta}(t,r) H^{-1/2}(\eps,\eta,r)U_{\eps,\eta}(r,t)\psi=H^{-1/2}(\eps,\eta,t)\psi+\cr
&+\int_t^r\di \tau\, U_{\eps,\eta}(t,\tau)  \frac{\partial }{\partial \tau}\(H^{-1/2}(\eps,\eta,\tau)\)U_{\eps,\eta}(\tau,t)\psi,
\end{align}
by using that $U_{\eps,\eta}(s,u)\Do(H_0)\subset \Do(H_0)$ for all $s,u\in\R$ and $\frac{\partial }{\partial \tau}\(H^{-1/2}(\eps,\eta,\tau)\)$ is a bounded operator, computed as follows.
By applying \cite[V-\S 3.11 equation (3.43)]{KatoChapter6} one has that
\begin{equation}
\label{eqn:H^{-1/2}(tau)}
H^{-1/2}(\eps,\eta,\tau)=\frac{2}{\pi} \int_0^{\infty} \di x\, {\(x^2+H(\eps,\eta,\tau)\)}^{-1},
\end{equation}
and thus
\begin{equation}
\label{eq: der H^-1/2}
\frac{\partial }{\partial \tau}H^{-1/2}(\eps,\eta,\tau)=-\frac{2\eps\eta g'(\eta \tau)}{\pi} \int_0^{\infty} \di x\, {\(x^2+H(\eps,\eta,\tau)\)}^{-1}  H_1{\( x^2+H(\eps,\eta,\tau) \)}^{-1}.
\end{equation}
Notice that in the above computation we have exchanged the derivative and the integral since by using condition~\eqref{Ch6eqn:cond on inf spect} and Lemma~\ref{Ch6lem:H_0^1/2 bounded implies H^1/2(t)bounded},  we obtain that 
\begin{align*}
& \abs{g'(\eta \tau)}\norm{{\(x^2+H(\eps,\eta,\tau)\)}^{-1}  H_1{\( x^2+H(\eps,\eta,\tau) \)}^{-1}}\leq\cr 
&\leq {M}^{\prime} \norm{{\( x^2+H(\eps,\eta,\tau) \)}^{-1}} \norm{H_1 H^{-1/2}(\eps,\eta,\tau)} \norm{ H^{1/2}(\eps,\eta,\tau) {\( x^2+H(\eps,\eta,\tau) \)}^{-1}}\cr
&\leq \frac{{M}^{\prime}}{1+x^2}(a+\eps M a^2)\qquad \text{for all $\tau\in\R$},
\end{align*}
where the right-hand term is integrable on $[0,\infty)$. Obviously, the previous bound implies that $\frac{\partial }{\partial \tau}H^{-1/2}(\eps,\eta,\tau)$ is bounded uniformly in time.
Moreover, notice that
\begin{equation}
\label{eqn: reg of derH^-1/2}
\frac{\partial }{\partial \tau}\(H^{-1/2}(\eps,\eta,\tau)\)\Do(H_0)\subset \Do(H_0).
\end{equation} 
Indeed for every $\phi\in\Do(H_0)=\Do(H(\eps,\eta,\tau))$ there exists $\varphi\in\Hi$ such that $\phi=H^{-1}(\eps,\eta,\tau)\varphi$ thus
\begin{align*}
\frac{\partial }{\partial \tau}H^{-1/2}(\eps,\eta,\tau)\phi&=-\frac{2\eps\eta g'(\eta \tau)}{\pi} \int_0^{\infty} \di x\, {\(x^2+H(\eps,\eta,\tau)\)}^{-1}  H_1H^{-1/2}(\eps,\eta,\tau)\cdot\cr
&\phantom{-\frac{2\eps\eta g'(\eta \tau)}{\pi} \int_0^{\infty} \di x\, } \cdot {\( x^2+H(\eps,\eta,\tau) \)}^{-1}H^{-1/2}(\eps,\eta,\tau)\varphi,
\end{align*}
by using condition~\eqref{Ch6eqn:cond on inf spect} and Lemma~\ref{Ch6lem:H_0^1/2 bounded implies H^1/2(t)bounded}, inclusion~\eqref{eqn: reg of derH^-1/2} is obtained. Therefore, we are allowed to apply $H^{1/2}(\eps,\eta,\tau)$ on the left-hand side of \eqref{Ch6eqn:en bound n=1 step I}, getting that for every $\psi\in\Do(H_0)$ 
\begin{align*}
&H^{1/2}(\eps,\eta,t)U_{\eps,\eta}(t,r) H^{-1/2}(\eps,\eta,r)\psi=U_{\eps,\eta}(t,r)\psi+\cr
&+\int_t^r\di \tau\, H^{1/2}(\eps,\eta,t) U_{\eps,\eta}(t,\tau)  \frac{\partial }{\partial \tau}\(H^{-1/2}(\eps,\eta,\tau)\)U_{\eps,\eta}(\tau,r)\psi.
\end{align*}
By multiplying the above equality on the left-hand side by $H^{-N/2}(\eps,\eta,t)$ and applying it to a particular subset of $\Do(H_0)\ni \psi=H^{N/2}(\eps,\eta, r)\phi$,
where $\phi\in\Do(H^{(N+2)/2}(\eps,\eta, r))$, we obtain that for every $\phi\in \Do(H^{(N+2)/2}(\eps,\eta, r))$
\begin{align}
\label{Ch6eqn:en bound n=N step I}
&H^{-N/2}(\eps,\eta,t)U_{\eps,\eta}(t,r)H^{N/2}(\eps,\eta,r)\phi=H^{-(N-1)/2}(\eps,\eta,t)U_{\eps,\eta}(t,r) H^{(N-1)/2}(\eps,\eta,r)\phi\cr
&\phantom{=}-\int_t^r\di \tau\, H^{-(N-1)/2}(\eps,\eta,t)U_{\eps,\eta}(t,\tau)  \frac{\partial }{\partial \tau}\(H^{-1/2}(\eps,\eta,\tau)\)U_{\eps,\eta}(\tau,r)H^{N/2}(\eps,\eta,r)\phi.
\end{align}
Therefore, in view of the induction hypothesis for $n=N-1$ we have that
\begin{align}
\label{Ch6eqn:en bound induction step I}
& \norm{H^{-N/2}(\eps,\eta,t)U_{\eps,\eta}(t,r)H^{N/2}(\eps,\eta,r)\phi}
\leq C_{N-1}(\eps)\norm{\phi}+\cr 
& +C_{N-1}(\eps)\int_t^r\!\!\di \tau\, \left\|H^{-(N-1)/2}(\eps,\eta,\tau)\frac{\partial }{\partial \tau}\(H^{-1/2}(\eps,\eta,\tau)\)H^{N/2}(\eps,\eta,\tau)\right.\cdot\cr
&\phantom{+C_{N-1}(\eps)\int_t^r\!\!\di \tau\,LL}\cdot\left.H^{-N/2}(\eps,\eta,\tau)U_{\eps,\eta}(\tau,r)H^{N/2}(\eps,\eta,r)\phi\right\|,
\end{align}
for $0\leq t\leq r\leq 1/\eta $.
Being $\Do(H^{(N+2)/2}(\eps,\eta, r))$ a core\footnote{
First of all, notice that ${\(\Id+\frac{1}{n}H^{(N+2)/2}(\eps,\eta, r)\)}^{-1}$ converges strongly to $\Id$. Indeed, in view of $\norm{{\(\Id+\frac{1}{n}H^{(N+2)/2}(\eps,\eta, r)\)}^{-1}}\leq 1$, if $v\in \Do(H^{(N+2)/2}(\eps,\eta, r))$ then 
\begin{align*}
\norm{{\(\Id+\frac{1}{n}H^{(N+2)/2}(\eps,\eta, r)\)}^{-1}v-v}&\leq\frac{1}{n}\norm{{\(\Id+\frac{1}{n}H^{(N+2)/2}(\eps,\eta, r)\)}^{-1}}\norm{H^{(N+2)/2}(\eps,\eta, r)v}\cr
&\leq\frac{1}{n}\norm{H^{(N+2)/2}(\eps,\eta, r)v}.
\end{align*}
By density of $\Do(H^{(N+2)/2}(\eps,\eta, r))$ in $\Hi$ the strong convergence follows. Therefore, for every $u\in\Do(H^{N/2}(\eps,\eta,r))$ defining $u_n:={\(\Id+\frac{1}{n}H^{(N+2)/2}(\eps,\eta, r)\)}^{-1}u\in \Do(H^{(N+2)/2}(\eps,\eta, r))$ one has that
$$
\lim_{n\to\infty}H^{N/2}(\eps,\eta,r)u_n=\lim_{n\to\infty}{\(\Id+\frac{1}{n}H^{(N+2)/2}(\eps,\eta, r)\)}^{-1}H^{N/2}(\eps,\eta,r)u=H^{N/2}(\eps,\eta,r)u,
$$
and thus by using that $H^{-N/2}(\eps,\eta,r)$ is bounded we obtain that $\lim_{n\to\infty}u_n=u$ as well.}  of $H^{N/2}(\eps,\eta,r)$, it suffices to prove the induction step on this set.
In order to conclude the proof, it is enough to observe that: 
For every $m\geq 1$, being $\alpha$, $\beta$ and $\gamma_m$ defined in \eqref{eqn: defn consts}, for all $\tau\in [0,1/\eta]$, for all $\psi\in \Do(H^{m/2}(\eps,\eta,\tau))$ , we have that
\begin{equation}
\label{Ch6eqn:bound to conclude}
\norm{H^{-(m-1)/2}(\eps,\eta,\tau)\frac{\partial }{\partial \tau}\(H^{-1/2}(\eps,\eta,\tau)\)H^{m/2}(\eps,\eta,\tau)\psi}\leq (\alpha +\beta \eps +\gamma_m)\eps\eta\norm{\psi}. 
\end{equation}
Indeed, notice that
\begin{align}
\label{eqn: final ineq}
&\norm{H^{-(m-1)/2}(\eps,\eta,\tau)\frac{\partial }{\partial \tau}\(H^{-1/2}(\eps,\eta,\tau)\)H^{m/2}(\eps,\eta,\tau)\psi}\cr
&\leq \norm{\frac{\partial }{\partial \tau}\(H^{-1/2}(\eps,\eta,\tau)\)H^{1/2}(\eps,\eta,\tau)\psi}+\cr
&\phantom{\leq}+\norm{\left[H^{-(m-1)/2}(\eps,\eta,\tau),\frac{\partial }{\partial \tau}\(H^{-1/2}(\eps,\eta,\tau)\)\right]H^{m/2}(\eps,\eta,\tau)\psi},
\end{align}
where each of the summands on the right-hand side is uniformly bounded in time as follows.
Being $\Do(H(\eps,\eta,\tau))$ a core of $\Do(H^{1/2}(\eps,\eta,\tau))$ \cite[V-\S 3.11 Lemma 3.38]{KatoChapter6}, in view of \eqref{eq: der H^-1/2}, above the first summand is bounded since for every $\widetilde{\psi}\in \Do(H(\eps,\eta,\tau))$
\begin{align*}
&\norm{\int_0^{\infty} \di x\, {\(x^2+H(\eps,\eta,\tau)\)}^{-1}  H_1{\( x^2+H(\eps,\eta,\tau) \)}^{-1}H^{1/2}(\eps,\eta,\tau)\widetilde{\psi}}\cr
&\leq\int_0^{\infty} \di x\, {\(x^2+1\)}^{-1} \norm{ H_1H^{-1/2}(\eps,\eta,\tau)}\norm{{\( x^2+H(\eps,\eta,\tau) \)}^{-1}H(\eps,\eta,\tau)\widetilde{\psi}}\cr
&\leq\frac{\pi}{2}(a+\eps M a^2)\norm{\widetilde{\psi}}.
\end{align*}
On the other hand for the second summand in \eqref{eqn: final ineq} for $m\geq 2$, we have that
\begin{align*}
&\left[H^{-(m-1)/2}(\eps,\eta,\tau),\frac{\partial }{\partial \tau}\(H^{-1/2}(\eps,\eta,\tau)\)\right]H^{m/2}(\eps,\eta,\tau)\psi\cr
&=\sum_{k=0}^{m-2}H^{-k/2}(\eps,\eta,\tau)\left[H^{-1/2}(\eps,\eta,\tau),\frac{\partial }{\partial \tau}\(H^{-1/2}(\eps,\eta,\tau)\)\right]H^{(k+2)/2}(\eps,\eta,\tau)\psi\cr
&=\frac{4\eps\eta\, g'(\eta \tau)}{\pi^2}\int_0^{\infty} \di x\int_0^{\infty} \di y\,{(x^2+H(\eps,\eta,\tau))}^{-1}{(y^2+H(\eps,\eta,\tau))}^{-1}\cdot\cr
&\phantom{=}\cdot\sum_{k=0}^{m-2}H^{-k/2}(\eps,\eta,\tau)[H(\eps,\eta,\tau),H_1]H^{(k-2)/2}(\eps,\eta,\tau)\cdot\cr
&\phantom{=}\cdot H(\eps,\eta,\tau){(x^2+H(\eps,\eta,\tau))}^{-1}H(\eps,\eta,\tau){(y^2+H(\eps,\eta,\tau))}^{-1}\psi.
\end{align*}
Clearly, the operator at right-hand side is uniformly bounded in $\tau$, since ${(x^2+H(\eps,\eta,\tau))}^{-1}$ and ${(y^2+H(\eps,\eta,\tau))}^{-1}$ ensure the uniform convergence of the integrals, 
 hypothesis~\ref{Ch6item:[H_1,H(eps,t)] H^-1 bounded} for $0 \leq k\leq m-2$ guarantees the boundedness of the middle factor and $\norm{ H(\eps,\eta,\tau){(z^2+H(\eps,\eta,\tau))}^{-1}}\leq 1$ for all $z\in [0,\infty)$.
Therefore, we obtain that
\[
\norm{\left[H^{-(m-1)/2}(\eps,\eta,\tau),\frac{\partial }{\partial \tau}\(H^{-1/2}(\eps,\eta,\tau)\)\right]H^{m/2}(\eps,\eta,\tau)\psi}\leq \eps\eta {M}^{\prime}\sum_{k=0}^{m-2}E_k\norm{\psi}.
\]
Finally, plugging estimate~\eqref{Ch6eqn:bound to conclude} into inequality~\eqref{Ch6eqn:en bound induction step I}, we have
\[
C_{\eps,\eta,N}(t,r)\leq C_{N-1}(\eps)\(1+(\alpha+\beta\eps +\gamma_N)\eps\eta  \int_t^r\di \tau\, C_{\eps,\eta,N}(\tau,r)\),
\]
for $0\leq t\leq r\leq 1/\eta $.
Applying Gr\"onwall's inequality, we conclude that 
\[
C_{\eps,\eta,N}(t,r)\leq C_{N-1}(\eps)\E^{ C_{N-1}(\eps)(\alpha+\beta\eps+\gamma_N)\eps\eta\abs{t-r}}\leq C_{N-1}(\eps)\E^{ C_{N-1}(\eps)(\alpha+\beta\eps+\gamma_N)\eps}=:C_N(\eps)
\]
for all $t,r\in [0,1/\eta]$. \qed

\subsection{Proof of Corollary~\ref{Ch6thm:en bounded scaled H}}
\label{ssec:proofofCor}
Notice that identity~\eqref{eqn:unitaries are related by the gauge transf} holds true since for every $\varphi\in\Do(H_0)$ one has that
\begin{align*}
&\iu \frac{\partial }{\partial s}\(\E^{\iu\frac{\eps}{\eta} \phi(s)H_1}U_{\eps,\eta}(s/\eta,u/\eta)\E^{-\iu \frac{\eps}{\eta}\phi(u)H_1}\varphi\)\cr
&=\E^{\iu\frac{\eps}{\eta} \phi(s)H_1}\(\frac{1}{\eta} H(\eps,\eta,s/\eta)-\frac{\eps}{\eta}g(s)H_1   \)U_{\eps,\eta}(s/\eta,u/\eta)\E^{-\iu \frac{\eps}{\eta}\phi(u)H_1}\varphi\cr
&=\frac{1}{\eta}\E^{\iu \frac{\eps}{\eta}\phi(s)H_1} H_0 \E^{-\iu \frac{\eps}{\eta}\phi(s)H_1}\E^{\iu \frac{\eps}{\eta}\phi(s)H_1} U_{\eps,\eta}(s/\eta,u/\eta)\E^{-\iu \frac{\eps}{\eta}\phi(u)H_1}\varphi=\frac{1}{\eta}\hat{H}(\eps,\eta,s)\hat{U}_{\eps,\eta}(s,u)\varphi,
\end{align*}
due to strong differentiability of $U_{\eps,\eta}(t,r)$ on $\Do(H_0)$, Assumption~\ref{domain power H0 invariant H1} for $k=2$ and $\Do(H_0)\subset \Do(H_1)$ by Assumption~\ref{Ch6item:H_1 H_0^1/2 bounded}, and similarly one verifies the other properties in \eqref{Ch6eqn: evol U hat}.
Therefore, fixed any $n\in\Np$, in view of Assumption~\ref{domain power H0 invariant H1} for $k=n$, for every $\psi\in \Do({H}^{n/2}_0)$ we have that
\begin{align*}
\hat{H}^{-n/2}(\eps,\eta,s)\hat{U}_{\eps,\eta}(s,u) \hat{H}^{n/2}(\eps,\eta,u)\psi&=\E^{\iu \frac{\eps}{\eta}\phi(s)H_1}H_0^{-n/2}\E^{-\iu \frac{\eps}{\eta}\phi(s)H_1}\E^{\iu\frac{\eps}{\eta} \phi(s)H_1}U_{\eps,\eta}(s/\eta, u / \eta) \cdot\cr
&\phantom{=}\cdot\E^{-\iu \frac{\eps}{\eta}\phi(u)H_1}\E^{\iu \frac{\eps}{\eta}\phi(u)H_1}H_0^{n/2}\E^{-\iu \frac{\eps}{\eta}\phi(u)H_1}\psi\cr
&=\E^{\iu\frac{\eps}{\eta} \phi(s)H_1}H_0^{-n/2}U_{\eps,\eta}(s/\eta, u / \eta) H_0^{n/2}\E^{-\iu \frac{\eps}{\eta}\phi(u)H_1}\psi.
\end{align*}
Thus, we deduce that
\begin{align*}
&\norm{\hat{H}^{-n/2}(\eps,\eta,s)\hat{U}_{\eps,\eta}(s,u) \hat{H}^{n/2}(\eps,\eta,u)\psi} \\
&\quad=\norm{H_0^{-n/2}H^{n/2}(\eps,\eta,s/\eta)\cdot H^{-n/2}(\eps,\eta,s/\eta)U_{\eps,\eta}(s/\eta, u / \eta)H^{n/2}(\eps,\eta,u/\eta)\cdot \right.\\
&\qquad \left. \cdot H^{-n/2}(\eps,\eta,u/\eta) H_0^{n/2}\E^{-\iu \frac{\eps}{\eta}\phi(u)H_1}\psi}\\
&\quad\leq C_n(\eps) (1+\eps D_n),
\end{align*}
by using Theorem~\ref{Ch6prop:en bounded} and Proposition~\ref{Ch6lem:bound prod opposite power H and H_eps}. Finally, the Riesz Lemma implies the thesis for all $n=-\abs{n}\in\Z$.
\qed

\section{Proof of the auxiliary results}
\label{sec:aux res}

\subsection{Proof of Lemma~\ref{Ch6lem:H_0^1/2 bounded implies H^1/2(t)bounded}}
In view of $\Do(H^{1/2}(\eps,\eta,t))=\Do(H^{1/2}_0)$ by Remark \ref{rem:aft ass}.\ref{item3:rem aft ass}, equality \eqref{eqn:H^{-1/2}(tau)} and the second resolvent identity, we have that
\begin{align}
\label{eqn: lem}
&H_1H^{-1/2}(\eps,\eta,t)=\frac{2}{\pi} \int_0^{\infty} \di x\, H_1{\(x^2+H(\eps,\eta,t)\)}^{-1}\cr
&=H_1H^{-1/2}_0-\frac{2\eps g(\eta t)}{\pi} \int_0^{\infty} \di x\,H_1 {\(x^2+H_0\)}^{-1}
 H_1 {\(x^2+H(\eps,\eta,t)\)}^{-1}.
\end{align}
In the last expression, for the second summand we observe that 
\begin{align*}
&\norm{\int_0^{\infty} \di x\,H_1 H^{-1/2}_0\cdot H^{1/2}_0{\(x^2+H_0\)}^{-1} H^{1/2}_0 \cdot H^{-1/2}_0H_1 \cdot {\(x^2+H(\eps,\eta,t)\)}^{-1}}\leq \frac{a^2\pi}{2},
\end{align*}
where we have used the hypothesis $\norm{H_1 H^{-1/2}_0}=a<\infty$, condition \eqref{Ch6eqn:cond on inf spect} and $\norm{H^{-1/2}_0H_1\varphi}= \norm{{\(H_1H^{-1/2}_0\)}^*\varphi}\leq a\norm{\varphi}$ for all $\varphi\in\Do(H_1)\supseteq \Do(H_0)$. Using the last inequality in \eqref{eqn: lem} the thesis is obtained.
\qed

\subsection{Proof of Proposition~\ref{Ch6lem:bound prod opposite power H and H_eps}}
First of all, notice that for any $k\in \Np$ if one supposes Assumption~\ref{Ch6item:power} then Remark~\ref{rem:aft ass}.\ref{item3:rem aft ass} ensures that the products of operators $H_0^{k/2}H^{-k/2}(\eps,\eta,t)$ and $H(\eps,\eta,t)^{k/2}H_0^{-k/2}$ are well defined on $\Hi$. 
We are going to prove inequality~\eqref{Ch6eqn:bound prod opposite power H and H_eps 1} for every $n\in\N$, proceeding by induction. The induction step will be proved by using the base cases for $0\leq n\leq 3$ and estimate \eqref{Ch6eqn:bound prod opposite power H and H_eps 2} for $n=1$.
For $n=0$ it is trivial.
For $n=1$, in view of equality~\eqref{eqn:H^{-1/2}(tau)} and the second resolvent identity we obtain that 
\begin{align*}
&\norm{H_0^{1/2}H^{-1/2}(\eps,\eta,t)}=\frac{2}{\pi}\norm{H_0^{1/2} \int_0^{\infty} \di x\, {\(x^2+H(\eps,\eta,t)\)}^{-1}}\cr
&\leq 1+ \frac{2\eps M}{\pi}\int_0^{\infty} \di x\norm{H_0^{1/2}{\(x^2+H_0\)}^{-1}H_0^{1/2}}\norm{H_0^{-1/2}H_1{\(x^2+H(\eps,\eta,t)\)}^{-1}}\cr
&\leq 1+ \eps M a,
\end{align*}
where we have used the hypothesis $\norm{H_1 H^{-1/2}_0}=a<\infty$ and condition \eqref{Ch6eqn:cond on inf spect}. Analogously, by virtue of Lemma~\ref{Ch6lem:H_0^1/2 bounded implies H^1/2(t)bounded} and condition \eqref{Ch6eqn:cond on inf spect}, one obtains \eqref{Ch6eqn:bound prod opposite power H and H_eps 2} for $n=1$.
For $n=2$ rewriting
\[
H_0 H^{-1}(\eps,\eta,t)\!=\!(H_0+\eps g(\eta t)H_1-\eps g(\eta t)H_1)H^{-1}(\eps,\eta,t)\!=\!\Id -\eps g(\eta t)H_1 H^{-1}(\eps,\eta,t),
\]
thus by applying Lemma~\ref{Ch6lem:H_0^1/2 bounded implies H^1/2(t)bounded} and condition \eqref{Ch6eqn:cond on inf spect}, inequality \eqref{Ch6eqn:bound prod opposite power H and H_eps 1} is obtained.
For $n=3$ notice that
\begin{align}
\label{eqn:ineq n=3}
&H_0^{3/2}H^{-3/2}(\eps,\eta,t)=H_0^{1/2}H_0H^{-1/2}(\eps,\eta,t)H^{-1}(\eps,\eta,t)=\cr
&H_0^{1/2}H^{-1/2}(\eps,\eta,t)H_0H^{-1}(\eps,\eta,t)+H_0^{1/2}[H_0,H^{-1/2}(\eps,\eta,t)]H^{-1}(\eps,\eta,t),
\end{align}
where on the right-hand side the first summand is bounded\footnote{In this proof when we write that an operator is bounded by a constant we mean it in the sense of the operator norm, and $\Or(\eps)$ is understood in the sense of the operator norm as well.} by $1+\Or(\eps)$ by applying the base cases for $1\leq n\leq 2$. For the second summand in \eqref{eqn:ineq n=3}, Leibniz's rule and equality~\eqref{eqn:H^{-1/2}(tau)} imply that
\begin{align*}
&H_0^{1/2}[H_0,H^{-1/2}(\eps,\eta,t)]H^{-1}(\eps,\eta,t)=\cr
&\frac{2\eps g(\eta t)}{\pi}\!\! \int_0^{\infty}\!\! \di x\, H_0^{1/2}{\(x^2+H(\eps,\eta,t)\)}^{-1}\cdot [H_1, H(\eps,\eta,t)]H^{-1}(\eps,\eta,t)\cdot{\(x^2+H(\eps,\eta,t)\)}^{-1}
\end{align*}
where in the last equality the first factor is uniformly bounded in $x$ since 
\begin{align*}
\norm{H_0^{1/2}{\(x^2+H(\eps,\eta,t)\)}^{-1}}&\leq\norm{H_0^{1/2}H^{-1/2}(\eps,\eta,t)H^{1/2}(\eps,\eta,t){\(x^2+H(\eps,\eta,t)\)}^{-1}}\cr
&\leq 1+A_1\eps,
\end{align*}
the second factor is bounded by virtue of hypothesis \ref{Ch6item:[H_1,H(eps,t)] H^-1 bounded} for $k=0$ and the last one ensures the convergence of the integral.
Now we take some $N\in\N$. We assume that inequality~\eqref{Ch6eqn:bound prod opposite power H and H_eps 1} holds true for $n\in\{1,\ldots,N-1\}$ and we prove it for $n=N$. We split the cases for even and odd $N$. 
Let $N=2m$ for $m\geq 2$, we get that
\begin{align}
\label{eqn: even case}
&H_0^{N/2}H^{-N/2}(\eps,\eta,t)=H_0^{m}H^{-m}(\eps,\eta,t)=\cr
&H_0^{m-1}[H_0,H^{1-m}(\eps,\eta,t)]H^{-1}(\eps,\eta,t)+H_0^{m-1}H^{1-m}(\eps,\eta,t)H_0 H^{-1}(\eps,\eta,t).
\end{align}
In \eqref{eqn: even case} the second summand is bounded by $1+\Or(\eps)$ by applying the induction hypothesis for $n=N-2$ and the base case for $n=2$. On the other hand, for the first summand in \eqref{eqn: even case} Leibniz's rule implies that
\begin{align*}
&H_0^{m-1}[H_0,H^{1-m}(\eps,\eta,t)]H^{-1}(\eps,\eta,t)=\cr
&\eps g(\eta t) H_0^{m-1} H^{1-m}(\eps,\eta,t) H^{-1}(\eps,\eta,t)\sum_{h=0}^{m-2}H^{m-h-1}(\eps,\eta,t)[H_1,H(\eps,\eta,t)]H^{h-m}(\eps,\eta,t),
\end{align*}
which is $\Or(\eps)$ thanks to the induction hypothesis for $n=N-2$, condition~\eqref{Ch6eqn:cond on inf spect} and hypothesis~\ref{Ch6item:[H_1,H(eps,t)] H^-1 bounded} for all $ 2-N\leq k:=2(h-m)+2 \leq -2$.
Let $N=2m+1$ for $m\geq 2$, similarly we have that
\begin{align*}
H_0^{N/2}H^{-N/2}(\eps,\eta,t)
&=H_0^{m-1/2}H_0 H^{-m+1}(\eps,\eta,t)H^{-3/2}(\eps,\eta,t)\\
&=H_0^{m-1/2}[H_0, H^{-m+1}(\eps,\eta,t)]H^{-3/2}(\eps,\eta,t)\\
&+H_0^{m-1/2}H^{1/2-m}(\eps,\eta,t)H^{1/2}(\eps,\eta,t) H_0^{-1/2}H_0^{3/2} H^{-3/2}(\eps,\eta,t),
\end{align*}
where in the last equality the second summand can be bounded by $1+\Or(\eps)$ due to the induction hypothesis for $n=N-2$, inequality \eqref{Ch6eqn:bound prod opposite power H and H_eps 2} for $n=1$ and the base case for $n=3$. While, the first summand can be rewritten as
\begin{align*}
&H_0^{m-1/2}[H_0, H^{-m+1}(\eps,\eta,t)]H^{-3/2}(\eps,\eta,t)\cr
&=\eps g(\eta t)H_0^{m-1/2}H^{-m+1/2}(\eps,\eta,t)\cdot\cr
&\phantom{=\eps g(\eta t)}\cdot H^{-1}(\eps,\eta,t)\sum_{h=0}^{m-2}H^{m-h-1/2}(\eps,\eta,t)[H_1,H(\eps,\eta, t)]H^{-m+h-1/2}(\eps,\eta,t),
\end{align*}
where last term is $\Or(\eps)$ in view of the induction hypothesis for $n=N-2$ and assumption \ref{Ch6item:[H_1,H(eps,t)] H^-1 bounded} for every $ 2-N\leq k:=2(h-m)+1 \leq -3$. Thus, inequality~\eqref{Ch6eqn:bound prod opposite power H and H_eps 1} is proved for every $n\in\N$. Similarly, one proves estimate~\eqref{Ch6eqn:bound prod opposite power H and H_eps 2} for all $n\in\N$.
Finally, to show inequality~\eqref{Ch6eqn:bound prod opposite power H and H_eps 1} for negative integer numbers, we notice that for any $n\in\Np$, for every $\psi\in\Do(H^{n/2}(\eps,\eta,t))$
\begin{align*}
\norm{H_0^{-n/2}H^{n/2}(\eps,\eta,t)\psi}= \norm{{\(H^{n/2}(\eps,\eta,t)H_0^{-n/2}\)}^*\psi}\leq \norm{H^{n/2}(\eps,\eta,t)H_0^{-n/2}}\norm{\psi},
\end{align*}
where the right-hand side is bounded by $(1+B_n\eps)\norm{\psi}$ in view of estimate~\eqref{Ch6eqn:bound prod opposite power H and H_eps 2} for positive integers. Analogously, one proves estimate~\eqref{Ch6eqn:bound prod opposite power H and H_eps 2} for negative integers.
\qed

\section{Application of the general strategy to Landau-type Hamiltonians}
\label{Ch6sec:appl} 
Among magnetic Schr\"odinger operators associated with non-interacting electrons in the plane, with (constant) magnetic field perpendicular to the plane,  the Landau model is emblematic for the understanding of the quantum Hall effect (QHE) \cite{Graf07}. For the model of Landau-type Hamiltonians an explanation for the QHE is provided \cite[Theorem 2.2]{ElgartSchlein04Chapter6} by relying on adiabatic perturbation theory \cite{Nenciu93}, which allows to compute rigorously the response of the intensity current being linear in the perturbation determined by the voltage difference (for recent topical reviews see \cite{HenheikTeufel21, MarcelliMonaco2}). Here, first we briefly explain why the energy estimates established in the general mathematical framework of Section \ref{sec: settingresults} are useful in this respect. Then, we verify that this model satisfies the assumptions previously stated.

This class of perturbed Hamiltonians is specified by \cite[equation $(1.1)$]{ElgartSchlein04Chapter6}. For the sake of clarity, we recall some definitions.

\begin{definition}
\label{Ch6defn:switch funct}
Let be $j\in\{1,2 \}$ and $l_j>0$, a \textbf{ $l_j$-switch function in the $j$-th direction} is a smooth function $\Lambda_j\colon\R^2\to [0,1]$ that depends only on the variable $x_j$ and satisfies
\[
\Lambda_j(x_j)=
\begin{cases}
0 & \text{if $x_j < -l_j$} \\
1 & \text{if $x_j > l_j$.}
\end{cases}
\]
\end{definition}

\noindent
We consider the unperturbed Hamiltonian $H_0$, defined as\footnote{We use Hartree atomic units, and moreover we reabsorb the factor $\frac{e}{c}$, where $e$ is the charge of the electron and $c$ is the speed of light, in the definition of the magnetic potential $\V{A}$.}
\begin{equation}
\label{Ch6eqn:unpert Landau H_0}
H_0:=\half \V{p}_{\V{A}}^2+\lambda V\quad\text{acting in $L^2(\R^2,\di \V{x})$,}
\end{equation}
where $\mathbf{p_A}:= (\mathbf{p}-\mathbf{A}(\mathbf{x}))$ with $\mathbf{p}:=-\iu \nabla=-\iu \left(\frac{\partial}{\partial x_1}, \frac{\partial}{\partial x_2} \right)$ and $\mathbf{A}(x_1,x_2):=B/2(-x_2,x_1)$ with $B>0$, $\lambda \in\R$ and the potential $V$ is such that $\norm{V}_{\infty}$ is finite\footnote{In \cite[Theorem 2.2]{ElgartSchlein04Chapter6} a stronger hypothesis is assumed, namely $\abs{\lambda}\norm{V}_{\infty}<B$ to ensure that the spectrum of $H_0$ consists of  a infinite sequence of bands, separated from each other by finite gaps.}.
\noindent 
The perturbed Hamiltonian is defined as\footnote{Notice that in this case we are imposing that the intensity of the perturbation and time-scale parameter, respectively $\eps$ and $\eta$, are equal.} $
H(\eps,t):=H(\eps,\eta=\eps,t)=H_0+\eps g(\eps t)\Lambda_1,$ where $0<\eps\ll 1$, $\Lambda_1$ is a $l_1$-switch function in the $1$-st direction and $g$ fulfills the hypotheses in Assumption~\ref{Ch6ass:perturbed H(t)}.
The multiplication operator $ \Lambda_1$ models an electric potential of negative unit drop for an electric field pointing in the negative $1$-st direction. One is interested in computing the Hall conductance $G\sub{Hall}$, defined as a ratio between the (excess of) induced current intensity when the perturbation is fully switched on and the voltage difference. More precisely, one introduces the operator $\iu [H_0,\Lambda_2]$ standing for the current intensity in the $2$-nd direction and $\rho_\eps(t)$ the density operator, representing the state of the system at time $t$ evolving from the Fermi projection $P_0$ of the unperturbed Hamiltonian $H_0$  with associated Fermi energy in a spectral gap of $H_0$. Thus, one is in shape to define the Hall conductance as
\begin{equation}
\label{eqn:defn GH}
\Tr\(\iu [H_0,\Lambda_2] (\rho_\eps(t)-P_0)\)=:-G\sub{Hall}\,\eps +o(\eps)\qquad \text{as $\eps\to 0$},
\end{equation}
for any $t\geq 1/\eps$ (when the perturbation is fully switched on).
In \cite{ElgartSchlein04Chapter6} first, by exploiting the invariance of the trace under unitary conjugation, one rewrites\footnote{The advantage of working with $\hat{H}(s)$ instead of $H(\eps,t)$ is the isospectrality of the former Hamiltonians.}
\begin{align}
\label{eqn:GH hatexp}
\Tr\(\iu [H_0,\Lambda_2] (\rho_\eps(s/\eps)-P_0)\)&=\Tr\(\E^{\iu \phi(s)\Lambda_1}\iu [H_0,\Lambda_2] (\rho_\eps(s/\eps)-P_0)\E^{-\iu \phi(s)\Lambda_1}\)\cr
&=\Tr\(\iu[\hat{H}(s),\Lambda_2](\hat{\rho}_\eps(s)-\hat{P}_0(s))\)
\end{align}
where $s:=\eps t$ is the scaled time, $\hat{H}(s):=\E^{\iu \phi(s)\Lambda_1} H_0 \E^{-\iu \phi(s)\Lambda_1} $, $\hat{\rho}_\eps(s):=\E^{\iu \phi(s)\Lambda_1} \rho_\eps(s/\eps)\E^{-\iu \phi(s)\Lambda_1}$ and $\hat{P}_0(s)=\E^{\iu \phi(s)\Lambda_1} P_0 \E^{-\iu \phi(s)\Lambda_1}$. Then, in order to derive an explicit formula for the Hall conductance $G\sub{Hall}$, they use an asymptotic expansion in $\eps$ powers of $\hat{\rho}_\eps(s)$
\begin{equation}
\label{eqn: asy hatrho}
\hat{\rho}_\eps(s)=\sum_{j=0}^{k}\eps^j B_j-\eps^{k}\int_0^s\di r\, \hat{U}_{\eps}(s,r)\dot{B}_k(r)\hat{U}_{\eps}(r,s),
\end{equation}
where $\hat{U}_{\eps}(r,s):=\hat{U}_{\eps,\eta=\eps}(r,s)$. Clearly, by plugging \eqref{eqn: asy hatrho} into \eqref{eqn:GH hatexp} and \eqref{eqn:defn GH}, beyond controlling the terms involving the $B_j$'s, one needs to estimate for $k>1$
\[
\eps^{k-1}\Tr\( \int_0^s\di r\, \iu [H_0,\Lambda_2]\hat{U}_{\eps}(s,r)\dot{B}_k(r)\hat{U}_{\eps}(r,s)\).
\]
In the continuum to prove that the trace of an operator $O$ is finite it suffices to show that $O$ has suitable localization both in energy and space \cite[Proposition 3.2]{ElgartSchlein04Chapter6}.
Since $\dot{B}_k(r)$ decays fast enough in energy in the sense of \eqref{eqn: enloc}, this energy localization is retained by the corresponding evolved operator $\hat{U}_{\eps}(s,r)\dot{B}_k(r)\hat{U}_{\eps}(r,s)$ as in \eqref{eqn: enlocevol} by exploiting the energy estimate in the form of \eqref{eqn:en est for H hat} (compare the inequality after \cite[equation $(3.12)$]{ElgartSchlein04Chapter6} and \cite[Remark (3), p. 599]{ElgartSchlein04Chapter6} for the case $n=0$). 

Now we are going to verify that the general assumptions of Section \ref{sec: settingresults} are fulfilled by this specific model.
Clearly, $H(\eps,t)$ satisfies Assumptions \ref{Ch6item:H_0 ass} and \ref{Ch6item:H_1 H_0^1/2 bounded}. Assumptions \ref{Ch6item:[H_1,H(eps,t)] H^-1 bounded}, \ref{Ch6item:power} and \ref{domain power H0 invariant H1} hold true under certain regularity conditions on $V$.  
Fix any $k\in\Z$, assume that the Sobolev norm\footnote{Let us recall that for $k\in\Np$ the Sobolev norm $\norm{f}_{k,\infty}$ of a scalar function $f$ on $\R^2$ is defined as
	$
	\norm{f}_{k,\infty}:=\sum_{\substack{\alpha_1,\alpha_2\in \Np \\ \alpha_1+\alpha_2\leq k}}\norm{\partial_{x_1}^{\alpha_1}\partial_{x_2}^{\alpha_2}f}_{\infty},
	$
	where $\norm{f}_{\infty}:=\sup_{\mathbf{x}\in\R^2}|f(\mathbf{x})|$. 
}$ \norm{V}_{\abs{k}+1,\infty}$ is finite then hypothesis~\ref{Ch6item:[H_1,H(eps,t)] H^-1 bounded} holds true. Indeed, since
$
[\Lambda_1,H(\eps,t)]=\frac{\iu}{2}\( p_{\mathbf{A},1}\Lambda'_1+\Lambda'_1  p_{\mathbf{A},1}\),
$
applying \cite[Proposition 3.1.(i)]{ElgartSchlein04Chapter6} we deduce that there exists a finite constant $e_k$:
\begin{align*}
\norm{H^{-k/2}(\eps,t)[\Lambda_1,H(\eps,t)]H^{(k-2)/2}(\eps,t)}\leq e_k \norm{\Lambda_1}_{\abs{k}+2,\infty},
\end{align*}
for all $\eps\in (0,1)$ and $t\in\R$. 

\noindent 
Now let $k\in\Np$ with $k\geq 2$, assume that $\norm{V}_{2(k-1),\infty}$ is finite then it follows that for all $\eps\in (0,1)$ and $t\in\R$
\[
\Do(H^{k}(\eps,t))\equiv\Do(H_0^k),
\]
namely the hypothesis~\ref{Ch6item:power} is fulfilled. Indeed, observe that 
\begin{equation}
\label{eqn:exp bin}
H^k(\eps,t)=H_0^k+{\(\eps g(\eps t)\)}^{k}\Lambda_1^k+\sum_{j=1}^{2^k-2}M_j,
\end{equation}
where each operator $M_j$ is such that there exist $\pmb{\alpha}=(\alpha_1,\dots,\alpha_k),\pmb{\beta}=(\beta_1,\dots,\beta_k)\in {\{0,1\}}^k$ with $\pmb{\alpha}\neq 0 \neq \pmb{\beta}$ and $\sum_{j=1}^k \alpha_j+\beta_j=k$:
\begin{equation*}
M_j={\(\eps g(\eps t)\)}^{\sum_{j=1}^k\beta_j}H_0^{\alpha_1}\Lambda_1^{\beta_1}\cdots  H_0^{\alpha_k}\Lambda_1^{\beta_k}.
\end{equation*}
We are going to show that $\Do(H^{k}_0)\subseteq \Do(H^{k}(\eps,t))$. It suffices to observe that every $M_j$ is densely defined on $\Do(H_0^{k-1})\supseteq \Do(H_0^k)$. In fact, rewriting\footnote{As in the previous sections, up to a shift of a constant, we can assume that $H_0\geq 1$.}
\begin{align*}
H_0^{\alpha_1}\Lambda_1^{\beta_1}\cdots H_0^{\alpha_k}\Lambda_1^{\beta_k}H_0^{-k+1}=&H_0^{\sum_{j=1}^{k}\alpha_j-k+1}\cdot\cr
&\cdot\prod_{m=1}^{k-1}H_0^{k-1-\sum_{j=0}^{m-1}\alpha_{k-j}}\Lambda_1^{\beta_{k-m}}H_0^{\sum_{j=0}^{m-1}\alpha_{k-j}-k+1}\cdot\cr
&\cdot H_0^{k-1}\Lambda_1^{\beta_k}H_0^{-k+1}
\end{align*}
here the product $\prod_{m=1}^{k-1}$ is ordered in the sense that a factor with larger index $m$ stands to the left of ones with smaller $m$ and, hence \cite[Proposition 3.1.(i).(b)]{ElgartSchlein04Chapter6} implies that
\begin{align*}
\norm{H_0^{\alpha_1}\Lambda_1^{\beta_1}\cdots H_0^{\alpha_k}\Lambda_1^{\beta_k}H_0^{-k+1}}\leq&C_{k-1}\norm{H_0^{\sum_{j=1}^{k}\alpha_j-k+1}}\norm{\Lambda_1^{\beta_k}}_{2k-2,\infty}\cdot\cr
&\cdot\prod_{m=1}^{k-1} C_{k-1-\sum_{j=0}^{m-1}\alpha_{k-j}}\norm{\Lambda_1^{\beta_{k-m}}}_{2k-2-\sum_{j=0}^{m-1}2\alpha_{k-j},\infty},
\end{align*}
which is finite, because $\sum_{j=1}^{k}\alpha_j-k+1\leq 0$ and any Sobolev norm of $\Lambda_1^{\beta_j}$ for all $\beta_j\in\{0,1\}$ is bounded. On the other hand, rewriting $H_0^k={\( H(\eps,t)-\eps g(\eps t)\Lambda_1 \)}^k$ and applying again \cite[Proposition 3.1.(i).(b)]{ElgartSchlein04Chapter6}, we deduce that $\Do(H_0^k)\supseteq \Do(H^k(\eps,t))$.
Now let $k\in\Np$, suppose that $\norm{V}_{k,\infty}$ is finite then Assumption~\ref{domain power H0 invariant H1} is satisfied. In fact, consider the gauge transformation $\E^{\iu  \lambda\Lambda_1}$ with $\lambda\in\R$, thus by virtue of \cite[Proposition 3.1.(i).(b)]{ElgartSchlein04Chapter6} we obtain that
$
\norm{H_0^{k/2}\E^{\iu  \lambda\Lambda_1}H_0^{-k/2}}\leq C_{k/2}\norm{\E^{\iu  \lambda\Lambda_1}}_{k,\infty}<\infty.
$
Thus, for every $n\in\Z$, if $\abs{n}\geq 2$ assuming that $\norm{V}_{\abs{n}-1,\infty}$ is finite, then  Theorem~\ref{Ch6prop:en bounded} implies that the inequality in \cite[Lemma 5.1]{ElgartSchlein04Chapter6} holds true. Furthermore, assuming that $\norm{V}_{2,\infty}$ is finite, then fixing any $n\in\Z$, if $\abs{n}\geq 2$ supposing in addition that $\norm{V}_{2\abs{n}-2,\infty}$ is finite one can apply Corollary~\ref{Ch6thm:en bounded scaled H} as well.


\bigskip

{\footnotesize

\begin{tabular}{ll}
(G. Marcelli)
             &  \textsc{Mathematics Area, SISSA} \\ 
        	&   Via Bonomea 265, 34136 Trieste, Italy \\
        	&  {E-mail address}: \href{mailto:giovanna.marcelli@sissa.it}{\texttt{giovanna.marcelli@sissa.it}}\\
\end{tabular}
} 

\end{document}